\documentclass[10pt,a4paper]{article}
\usepackage{geometry}
\usepackage{graphicx}
\usepackage{amsmath}
\usepackage{listings}
\usepackage{framed}
\usepackage{xcolor}
\usepackage{authblk}
\DeclareGraphicsExtensions{.pdf}
\newenvironment{cframed}[1][blue]
  {\def\FrameCommand{\fboxsep=\FrameSep\fcolorbox{#1}{white}}%
    \MakeFramed {\advance\hsize-\width \FrameRestore}}
  {\endMakeFramed}
\author[1]{Dariush Hassani \thanks{dariush.hassani@outlook.com}}
\author[2]{Shahnoosh Rafibakhsh \thanks{Corresponding author: rafibakhsh@srbiau.ac.ir}}
\affil[1,2]{Department of Physics, Science and Research Branch, Islamic Azad University, Tehran, Iran}
{
    \makeatletter
    \renewcommand\AB@affilsepx{: \protect\Affilfont}
    \makeatother

}
\title{Parallelization and implementation of multi-spin Monte Carlo simulation of 2D square Ising model using MPI and C++}
\date{}

\begin{document}
\maketitle
\begin{abstract}
In this paper, we present a parallel algorithm for Monte Carlo simulation of the 2D Ising Model to perform efficiently on a cluster computer using MPI. We use C++ programming language to implement the algorithm. In our algorithm, every process creates a sub-lattice and the energy is calculated after each Monte Carlo iteration. Each process communicates with its two neighbor processes during the job and they exchange the boundary spin variables. Finally, the total energy of lattice is calculated by map-reduce method versus the temperature. We use multi-spin coding technique to reduce the interprocess communications. This algorithm has been designed in a way that an appropriate load-balancing exists and it benefits a good scalability. It has been executed on the cluster computer of Plasma Physics Research Center which includes 9 nodes and each node consists of two quad-core CPUs. Our results show that this algorithm is more efficient for large lattices and more iterations.
\end{abstract}

\section{Introduction} \label{sect.intro}
The Ising model \cite{Ising1925} gives a microscopic description of the ferromagnetism which is caused by the interaction between spins of the electrons in a crystal. The particles are assumed to be fixed on the sites of the lattice. Spin is considered as a scalar quantity which can achieve two values $+1$ and $-1$. The model is a simple statistical one which shows the phase transition between high temperature paramagnetism phase and low temperature ferromagnetic one at a specific temperature. In fact, the symmetry between up and down is spontaneously broken when the temperature goes below the critical temperature.  However, the one-dimensional Ising model, which has been exactly solved- shows no phase transition. The two-dimensional Ising model has been solved analytically with zero \cite{onsager} and non-zero \cite{baxter} external field. In spite of a lot of attempts to solve 3D Ising model, one might say that this model has never been solved exactly. All the results for the three-dimensional Ising model have been used approximation approaches and Monte Carlo methods.

Monte Carlo methods or statistical simulation methods are widely used in different fields of science such as physics, chemistry, biology, computational finance and even new fields like econophysics \cite{PhysRevB.84.094431, KOZUBSKI2010S80, LYBERATOS2016266, MASROUR201353, PhysRevB.72.094203, YANG2005417}. The simulation can proceed by sampling from the Probability Density Function (PDF) and generating random numbers uniformly. The simulation of the Ising model on big lattices increases the cost of simulation. One way to reduce the simulation cost is to design the algorithms which work faster. Swendsen-Wang and Wolff algorithms \cite{Swendsen:1987ce,Wolff:1988uh} and multi-spin coding methods \cite{jacobs1981, Williams1984, ZORN1981337} are the examples of such methods. Another way is to parallelize and execute the model on GPU’s, GPU clusters and cluster computers \cite{block2010, block2012, hawick2011, komura2012a, komura2012b, preis2009, komura2014, ALTEVOGT19931041, ito1993, Wansleben1984, barkema1994, PhysRevE.81.026701, WEIGEL20111833}.

In this paper, we present a parallel algorithm to simulate the 2D Ising model using Monte Carlo Method. Then, we run the algorithm on a cluster computer using C++ programming language and MPI. Message Passing Interface (MPI) is a useful programming model in HPC systems \cite{petrov2014,geng2013,keppens2012,oger2016,cheng2017,leboeuf2016,wang2015} ‬‬in which the processes communicate through message-passing and was designed for distributed memory architectures. MPI provides functionalities which allow two specified processes to exchange data by sending and receiving messages. To get high efficiency, it is necessary to have good load balancing and also to have minimum communications between processes.

In our algorithm, each individual process creates its own sub-lattice, initializes it, gets all Monte Carlo iterations done and calculates the energy of the sub-lattice for a specific temperature. Each process communicates with its two neighbor processes during the job and they exchange the boundary spin variables. Finally, the total energy of lattice is calculated by map-reduce method. Since in multi-spin coding technique each spin is stored by 3 bits, interprocesses communications are reduced considerably. Because computational load of each sub-lattice is assigned to each process and size of all sub-lattices is equal, an appropriate load-balancing exists. Since each process - independent of number of processes - only communicates with its two neighbor processes and the lattice is decomposed into sub-lattices, the algorithm benefits a good scalability.

This paper has been organized as follows. In section \ref{sect.ising}, Metropolis algorithm and the Ising model are studied briefly. In section \ref{sect.multispin}, we explain how to use Multi-spin coding method to calculate the interaction energy between a specific spin and its nearest neighbors. We also study the boundary conditions in the memory-word lattice. Details of parallelization of the algorithm is discussed in section \ref{sect.parallel} and the method of implementation is given in section \ref{sect.implementation}. Finally, the results are given in section \ref{sect.result}.

\section{Metropolise algorithm and Ising model} \label{sect.ising}

The Ising model consists of spins variables which take values $+1$ or $-1$ and are arranged in a one, two or three dimensional lattice. Each spin interacts with its neighbors and the interaction is given by the Hamiltonian:
\begin{equation}\label{eq.Hamiltonian}
  {\cal H}=-J\sum_{<m,n>}^{} s_m s_n,
\end{equation}
where $J$ is the coupling coefficient. The summation in Eq.~\eqref{eq.Hamiltonian} is taken over the nearest neighbor pairs $<m,n>$. Periodic boundary conditions are used which state that spins on one edge of the lattice are neighbors with the spins on the opposite side. In this paper, we focus on simulation of the 2D square Ising model using Metropolis Monte Carlo algorithm \cite{Metropolis1953}. The lattice is initialized randomly and is updated as the following:
\begin{enumerate}
\item Select a spin ($s_{i,j}$) randomly and calculate the interaction energy between this spin and its nearest neighbors ($E$).
\item Flip the spin $s_{i,j}$ to $s^{\prime}_{i,j}$ and again calculate the interaction energy ($E^{\prime}$).
\item $\triangle E=E^{\prime}-E$, if $\triangle E\leq 0$, $s^{\prime}_{i,j}$ is accepted. Otherwise, $s^{\prime}_{i,j}$ is accepted with the probability $e^{-\triangle E/{KT}}$ where $K$ is Boltzmann constant and $T$ is the temperature.
\item Repeat steps $1-3$ till we are sure that every spin has been flipped.
\item Calculate theh total energy of the lattice for $i^{th}$ iteration $(E_{total}^i)$.
\end{enumerate}
The steps above form a Monte Carlo iteration. We perform enough iterations ($N$ times) and finally average on $(E_{total}^i)$ to obtain $E_{total}$:
\begin{equation}\label{eq.Eave}
  E_{total}=\frac{1}{N}\sum_{i=1}^{N}E_{total}^i.
\end{equation}

\section{Multi-spin coding method} \label{sect.multispin}

Multi-spin coding refers to all techniques that store and process multiple spins in one memory word. In this paper, we apply the multi-spin coding technique to the 2D Ising model. In general, multi-spin coding technique results in a faster algorithm as a consequence of updating multiple spins simultaneously. However, we mainly employ this technique to reduce the interprocess communications.

\begin{figure}
  \centering
   \includegraphics[width=0.7\linewidth]{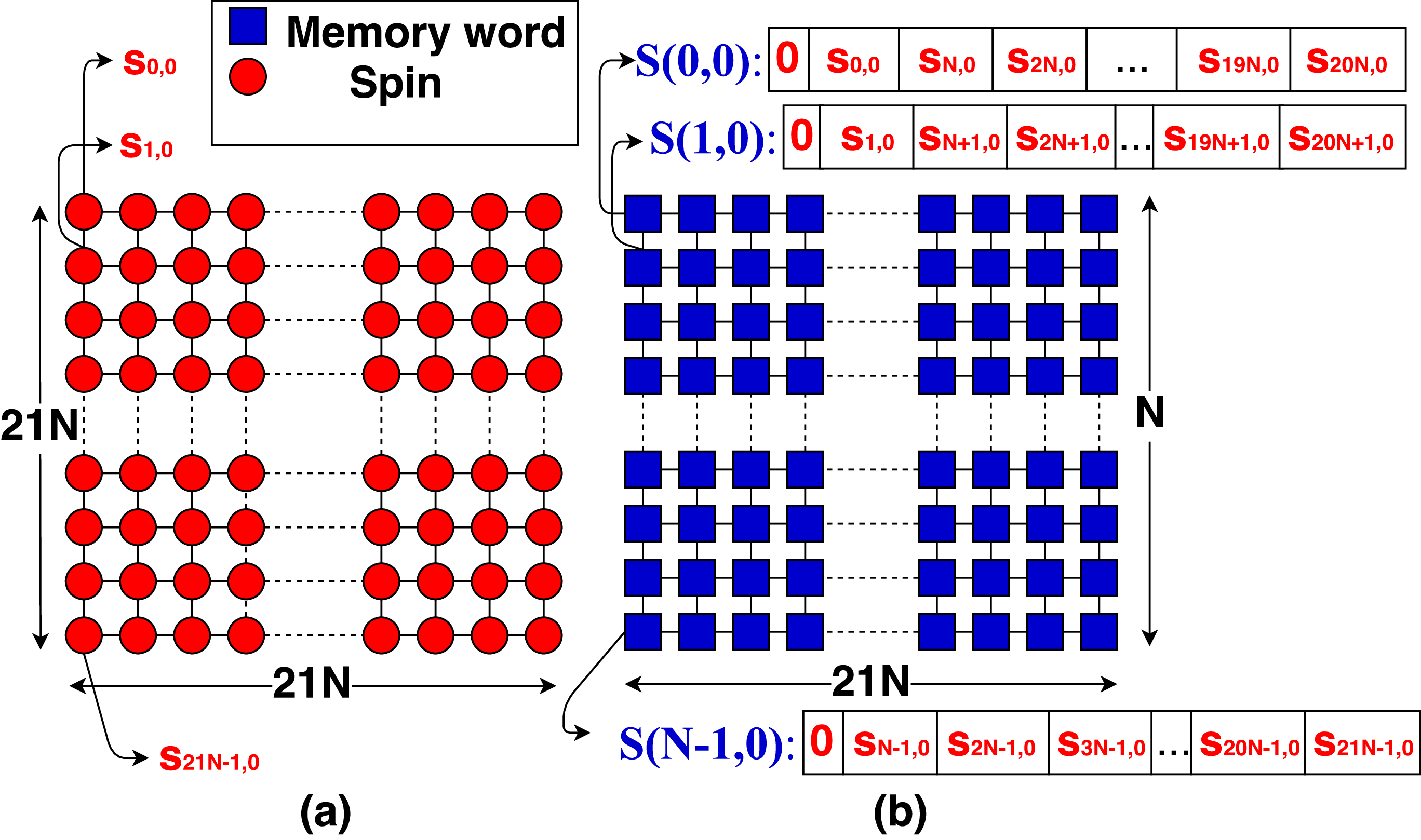}
  \caption{Arranging spins in memory words. (a) Spin lattice  (b) Memory-word lattice}\label{fig.multispin}
\end{figure}

We apply the multi-spin coding introduced in ref.~\cite{jacobs1981}. However, in our implementation, the size of a memory word is $64$ bits, in contrast to Jacobs's $60$-bit memory word. In addition, each spin is retained in three consecutive bits and the value of the $64^{th}$ bit is always set to zero. $000$ represents the spin down and the spin up is shown by $001$. Since a memory word contains $21$ spins, the size of the lattice is taken to be $21N \times 21N$, where $N$ is an integer greater than one. Now, we need to convert the spin lattice (Fig.~\ref{fig.multispin}.a) to the lattice of memory words (Fig.~\ref{fig.multispin}.b). Therefore, the size of the memory word lattice is considered as $N \times 21N$.
Each column of the spin lattice is coded into the same column of the memory word in the memory word lattice. So, $21N$ spins in one column of the spin lattice are arranged in $N$ memory words of a column in the memory word lattice as follows:
\begin{flalign*}
  &S(0,J):&&s_{0,j}, s_{N,j}, s_{2N,j}, ... , s_{19N,j}, s_{20N,j}&\\
  &S(1,J):&&s_{1,j}, s_{N+1,j}, s_{2N+1,j}, ... , s_{19N+1,j}, s_{20N+1,j}&\\
  &\vdots &&\vdots &\\
  &S(N-1,J):&&s_{N-1,j}, s_{2N-1,j}, s_{3N-1,j}, ... , s_{20N-1,j}, s_{21N-1,j},&
\end{flalign*}
where $S(I,J)$ represents the memory word at the row $I$ and the column $J$, $0 \leq I \leq N-1$, $0 \leq J \leq 21N-1$. $s_{i,j}$ shows the spin located at the row $i$ and the column $j$ where $j=J$. The advantage of this arrangement is that each spin is placed in the appropriate position related to its neighbors. Consider $k^{th}$ spin in a given memory word $S(I,J)$. The right/left/top/down neighbor of the $k^{th}$ in the spin lattice is exactly $k^{th}$ spin in the right/left/top/down neighbor of the memory word $S(I,J)$ in the memory word lattice.

\begin{figure}
  \centering
  \includegraphics[width=0.7\linewidth]{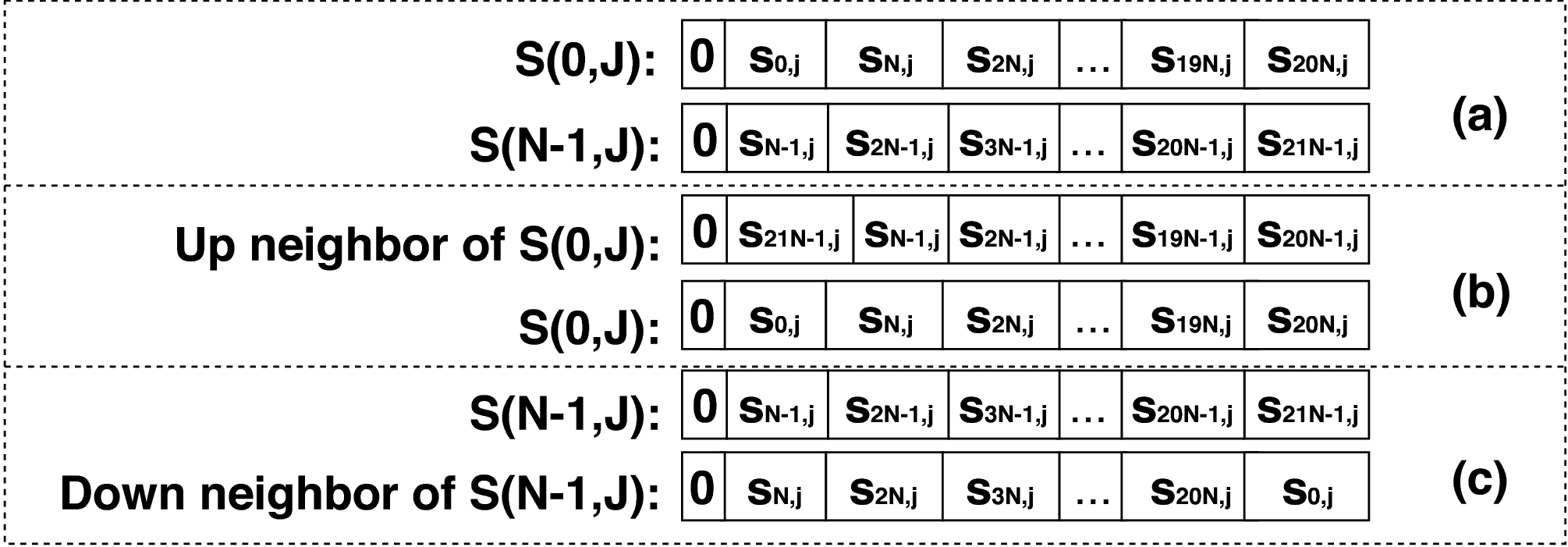}
  \caption{(a) Memory words in the first and last rows of a memory word lattice (b) up neighbor of $S(0,J)$ formed by shifting three bits of $S(N-1,J)$ to the right (c) down neighbor of $S(N-1,J)$ formed by shifting three bits of $S(0,J)$ to the left.}\label{fig.boundary}
\end{figure}

In order to apply periodic boundary conditions to the memory words in the first and last row (Fig.~\ref{fig.boundary}.a), we need to make some changes to up and down neighbors in advance. In fact, the down (up) neighbor of $S(N-1,J)$ ($S(0,J)$) is not exactly $S(0,J)$ ($S(N-1,J)$). For a memory word in the first (last) row, its up (down) neighbor -which is the memory word in the last (first) row and in same column- have to be shifted 3 bits to the right (left). These two cases have been shown in the diagrams (b) and (c) of Fig.~\ref{fig.boundary}.
We should recall that the $64^{th}$ bit is always set to zero.

\subsection{Calculation of energy}

Now, using multi-spin coding method, we show how to calculate the energy difference ($\Delta E$) between two configurations in the Ising model. At first, to better understand the process, we consider two 3bit-spins $s_1$ and $s_2$. $s_1\  XOR\  s_2$ produces $000$ when the two spins are placed in the same direction and $001$ is given when spins $s_1$ and $s_2$ are in the opposite directions. Hence, for a given memory word $S(I,J)$, the expression
\begin{eqnarray}\label{eq.xor}
\begin{aligned}
   &(S(I,J)\  XOR\  S(I-1,J)) + (S(I,J)\  XOR\  S(I+1,J))+\\
   &(S(I,J)\  XOR\  S(I,J-1)) + (S(I,J)\  XOR\  S(I,J+1)),
\end{aligned}
\end{eqnarray}
generates a value in the range of $[0,4]$ for every 3bit-group given in the last column of Table~\ref{table.multi-spin}. In the second and third columns, we have considered different cases that might occur between a selected spin and its four neighbors. The initial interaction energy $E$ and the energy $E^\prime$ calculated after flipping the selected spin have been presented in the forth and fifth rows, respectively.
\begin{table}
  
   \begin{tabular}{|c|c|c|c|c|c|c|}
    \hline
      Configuration & Selected Spin & Nearest Neighbors & $E$ & $E^\prime$ & $\Delta E$ & Value of a 3-bit group \\
    \hline
    \hline
    1 & Up & 4 Up - 0 Down & -4J & 4J & 8J & 000 \\
    \cline{1-5}
    2 & Down & 0 Up - 4 Down & -4J & 4J &  &  \\
    \hline
    3 & Up & 3 Up - 1 Down & -2J & 2J & 4J & 001 \\
    \cline{1-5}
    4 & Down & 1 Up - 3 Down & -2J & 2J &  &  \\
    \hline
    5 & Up & 2 Up - 2 Down & 0 & 0 & 0 & 010 \\
    \cline{1-5}
    6 & Down & 2 Up - 2 Down & 0 & 0 &  &  \\
    \hline
    7 & Up & 1 Up - 3 Down & 2J & -2J & -4J & 011 \\
    \cline{1-5}
    8 & Down & 3 Up - 1 Down & 2J & -2J &  &  \\
    \hline
    9 & Up & 0 Up - 4 Down & 4J & -4J & -8J & 100 \\
    \cline{1-5}
    10 & Down & 4 Up - 0 Down & 4J & -4J &  &  \\
    \hline
  \end{tabular}
  \caption{Different configurations which might happen between a selected spin and its four nearest neighbors. The initial energy $E$ and the energy $E^\prime$ after flipping the selected spin, have been shown in forth and fifth columns, respectively. The energy difference $\Delta E=E^\prime-E$ has been given in the sixth column. the last column represents the value of a 3-bit group. \label{table.multi-spin}}
\end{table}

\section{Parallelization} \label{sect.parallel}

\begin{figure}
  \centering
  \includegraphics[width=0.7\linewidth]{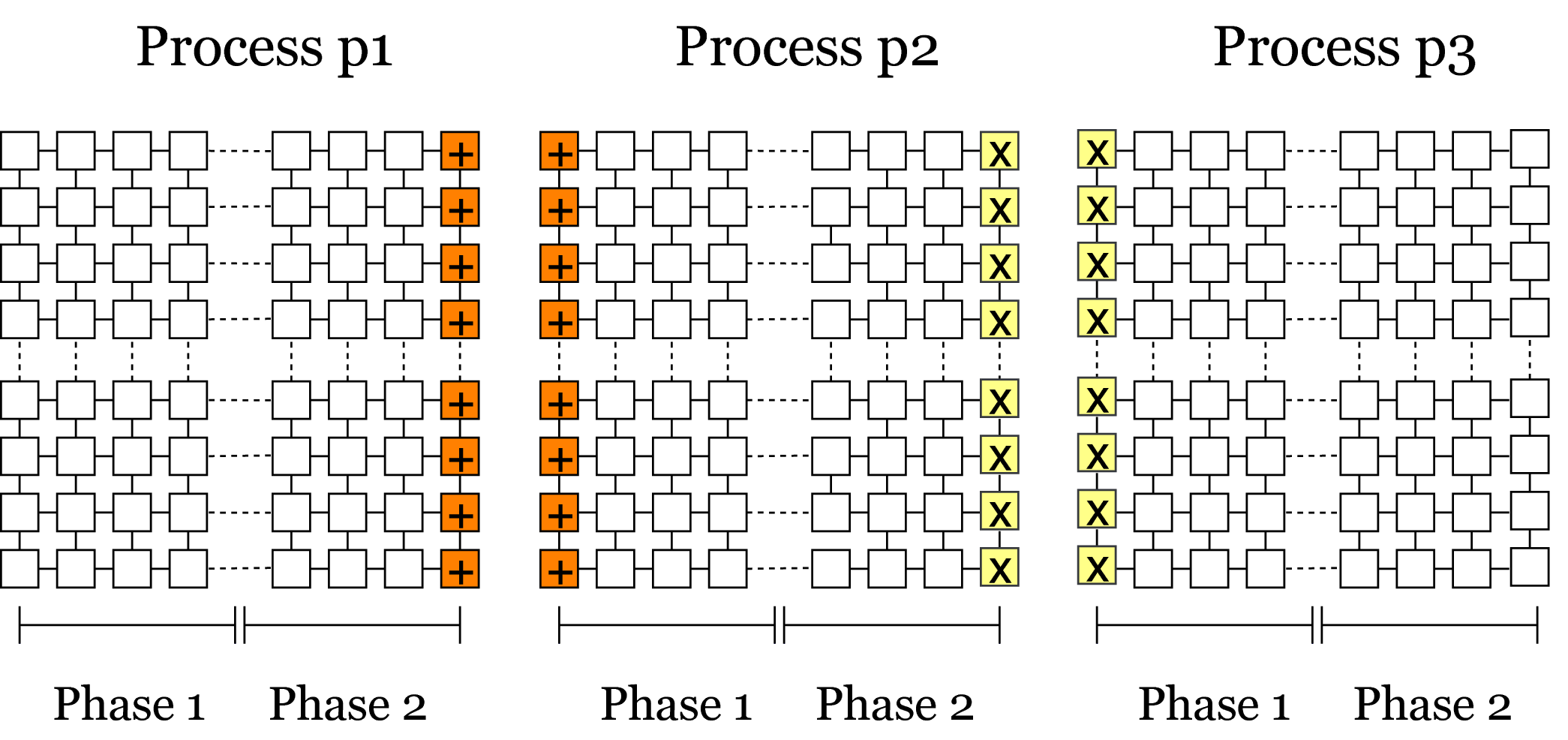}
  \caption{Sub-lattices of the memory words that belong to the three consecutive processes. The memory words on the border of the two different sub-lattices which have interaction with each other, are marked the same. Each half of the memory word lattice is updated in the phase 1 or 2. Therefore, the border memory words are not updated simultaneously.}\label{fig.sublattices}
\end{figure}

In a Monte Carlo Metropolis iteration, each memory word is updated at least once. The iterations must be performed enough times to yield accurate outcome energy. The given lattice could be vertically divided into $N_p$ sub-lattices with equal sizes, where $N_p$ is the number of processes. Computational load of each sub-lattice is assigned to the processes $0$ to $N_p-1$ from left to right. Each process creates a sub-lattice of the specific size, initializes the sub-lattice, performs all Monte Carlo iterations and calculates the energy of the sub-lattice using Eq.~\eqref{eq.Eave}. When all individual processes calculate the energy of their own sub-lattice, the energies of the sub-lattices are added up, through a Map-Reduce operation, to calculate the total energy of the lattice . However, this approach results in two problems. As illustrated in Fig.~\ref{fig.sublattices}, half of the neighbors of the memory words on the border, are placed in the sub-lattice of neighbor process. Therefore, to calculate the energy of the memory words on the border, some memory words of the side sub-lattice are needed. Therefore, these memory words have to be observed when needed. Moreover, we should note that neighbor memory words should not be updated simultaneously by different processes.

To deal with the second problem, we propose a method in which the memory words are updated in two phases (see Fig.~\ref{fig.sublattices}). In each phase, half of the sub-lattice is updated while the other half stays unchanged i.e. in phase 1 (2) the left (right) half of the sub-lattice is updated. After the phase 1 (2) is done, each process will pass to phase 2 (1) only if its right (left) process accomplishes the phase 1 (2). To better understand this process, we consider three consecutive processes in Fig.~\ref{fig.sublattices} which are updating the left half of their sub-lattices in phase 1. The process p2 has updated the left half of its sub-lattice and is going to start the phase 2 to update the right half. However, it is not able to reach the phase 2 until the process p3 accomplishes the phase 1 and finishes the update of the left half. So, the memory words on the borders of the processes p2 and p3, marked with $\times$ in Fig.~\ref{fig.sublattices}, do not update simultaneously. In the same manner, the memory words on the borders of processes p1 and p2, marked with + in Fig.~\ref{fig.sublattices} do not update at the same time.

Now, we turn to the first problem. As mentioned before, in each phase half of a sub-lattice is updated. Before a process starts updating the half of the sub-lattice, it should receive the corresponding border memory words of the neighbor process. Suppose that the process p2 is going to update the left half of its sub-lattice in phase 1. It waits to receive the right-side border memory words of the process p1. The process p1 sends its right border memory words to the process p2 asynchronously just after it accomplishes the phase 2 of the last iteration. After p2 receives the border memory words from p1 synchronously, it starts updating the memory words in the phase 1. Just after finishing the phase 1, p2 sends its updated left-side border memory words to p1 asynchronously and goes to the phase 2. The similar procedure occurs for other processes as well. It should be mentioned that we use periodic boundary conditions thereby the left neighbor of the first process is the last process, and likewise the right neighbor of the last process is the first process.

\section{Implementation} \label{sect.implementation}
\begin{figure}
  \centering
  \includegraphics[width=0.7\linewidth]{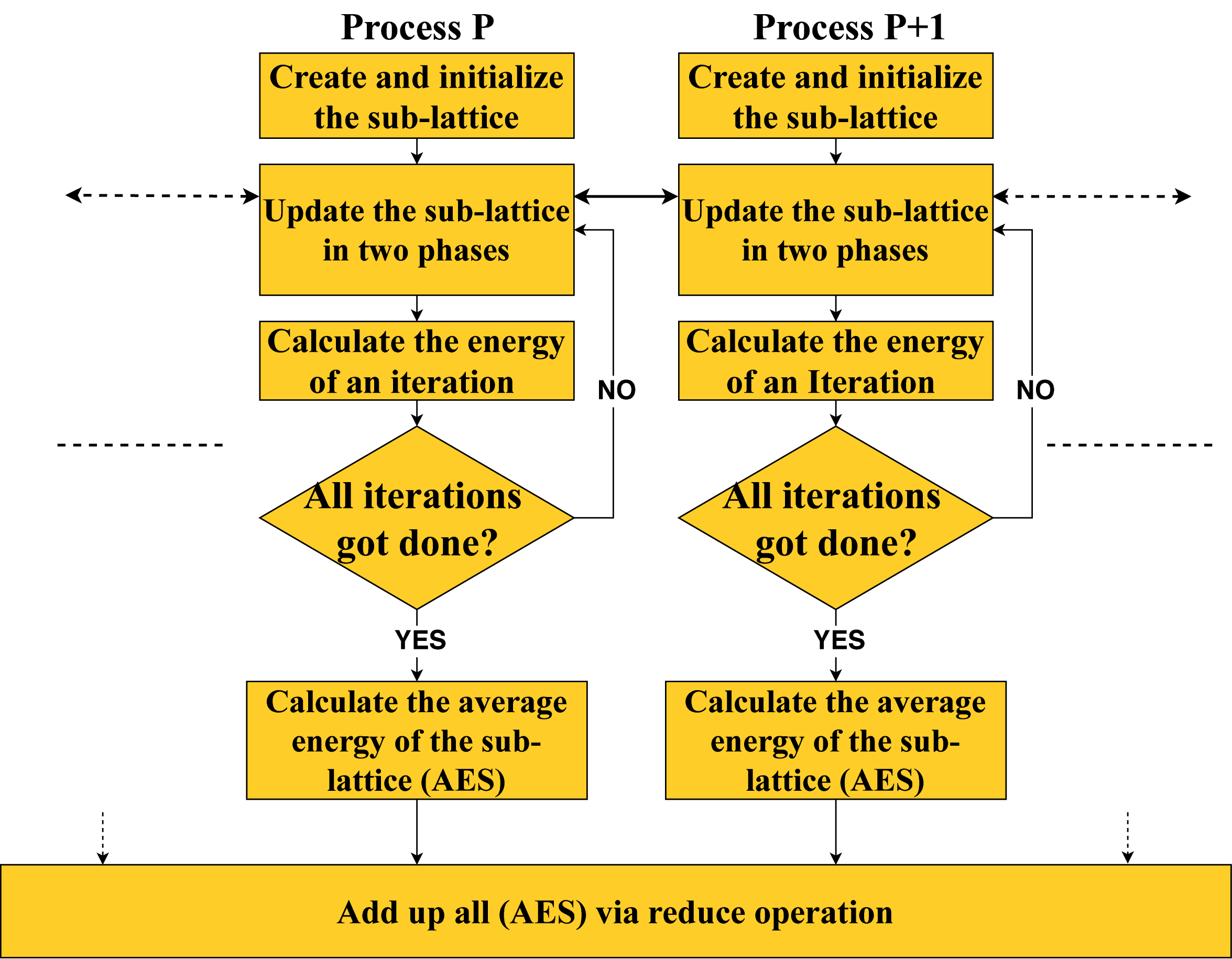}
  \caption{Function of our implemented program. Left and right arrows denote interprocess communications.}\label{fig.flowchart}
\end{figure}

In this section, we describe the implementation details of the algorithm presented in the previous section. Consider a memory word lattice of size $N\times 21N$ where $N$ is an arbitrary integer bigger than one. We execute the algorithm on $N_p$ processes and each process is identified by an integer number, $0$ to $N_p -1$, called rank. Each process is responsible for $N_c$ columns of the memory word lattice where $N_c=\frac{21N}{N_p}$. Each individual process creates its own sub-lattice, initializes it, gets all Monte Carlo iterations done and calculates the energy of the sub-lattice for a specific temperature. Within each Monte Carlo iteration a sub-lattice is updated many times and the energy of iteration is calculated. Finally, the total energy of the memory word lattice for a specific temperature is obtained via a reduce operation. This operation is illustrated in Fig.~\ref{fig.flowchart}.
Now, every step of the algorithm is studied in details.

\subsection{Initialization} \label{sub.sect.Init}

\begin{figure}
  \centering
  \includegraphics[width=0.7\linewidth]{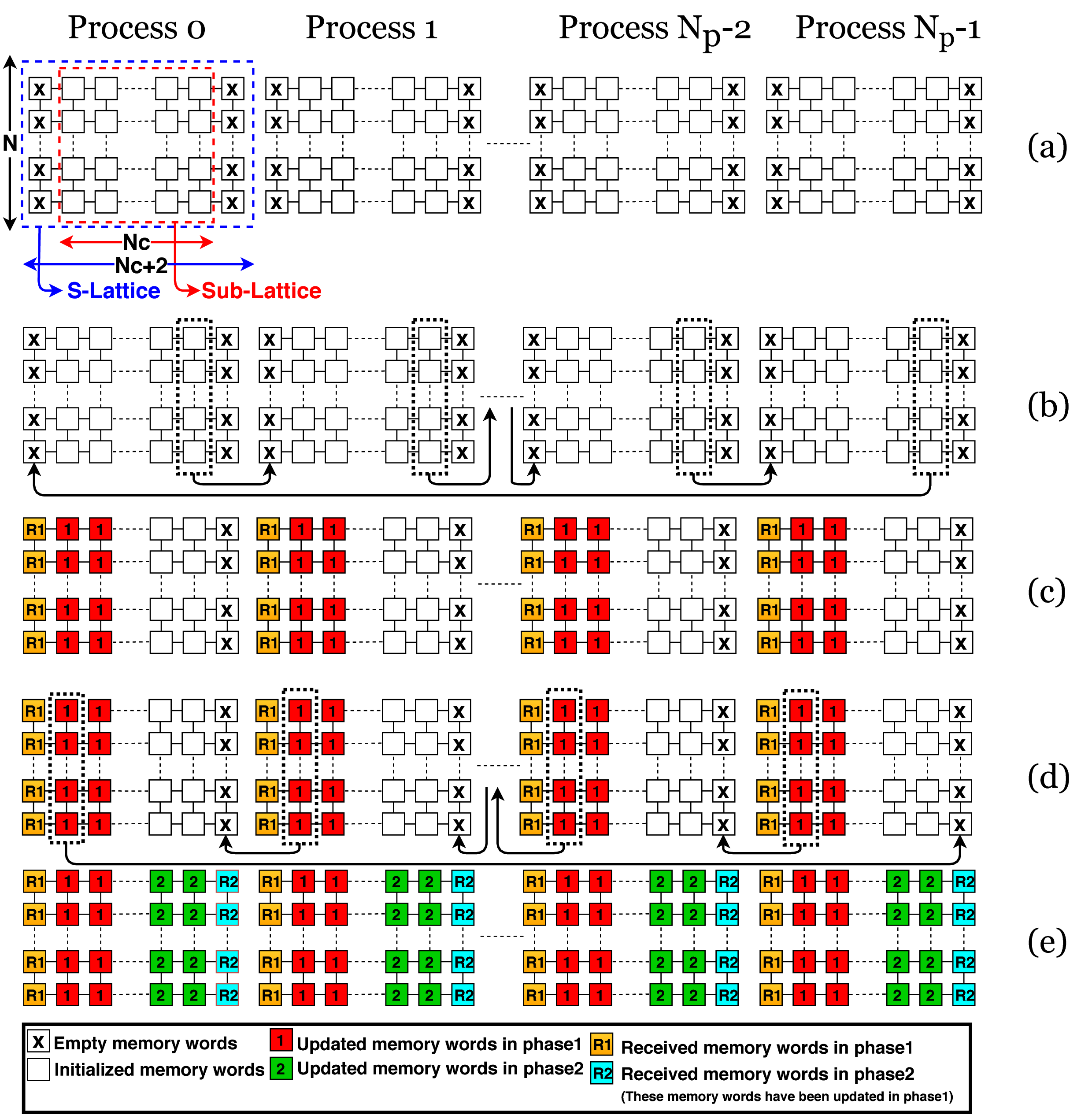}
  \caption{(a) S-lattice of each process. (b) Communication between processes before updating the memory words in phase 1. (c) Updating memory words in phase 1. (d) Communication between processes before updating memory words in phase 2. (e) Updating the memory words in phase 2. }\label{fig.updating}
\end{figure}

Now, we explain how a process creates its own sub-lattice and initializes it randomly. We use a 64-bit long integer as a memory word to store $21$ spins. A 2D array of $N\times N_c$ long integers forms the sub-lattice of the process where $N_c=\frac{21N}{N_p}$. However, we use an array with two additional columns ($N\times (N_c+2)$) reserved for border memory words of the neighbor processes (see Fig.~\ref{fig.updating}.a).
In this paper, we denote this array by S-lattice. Each spin in the memory words of the s-lattice is initialized with $0$ or $1$ randomly -except for the first and last columns- which represent spin down and up, respectively.

\subsection{Updating} \label{sub.sect.Update}
As mentioned before, updating process is done in two phases. In each phase, one half of the sub-lattice is updated. In cases where $N_c$ is odd, $floor(N_c/2)$ columns are updated in phase 1 and the rest of the columns is updated in phase 2. Before starting the update process in each phase, some interprocess communication should be carried out.

At first, each process sends its border memory words to its neighbor process asynchronously. Then, it waits to receive the border memory words from its neighbor processes. When the process receives the required border memory words, it can accomplish the phase by frequently updating the memory words that belong to the corresponding phase. In phase 1, in which the left half of the sub-lattice is updated, each process sends the rightmost column of its sub-lattice to its right neighbor (Fig.~\ref{fig.updating}.b). So, the destination of the sending memory words is determined by the following code:
\begin{verbatim}
destination = (Rank == NP-1) ? 0: Rank+1;
\end{verbatim}
It means that, sue to the periodic boundary conditions, the right neighbor of the process with the rank $N_p-1$ is the process $0$. Likewise, the source process from which the process receives the border memory words is determined by the following code:
\begin{verbatim}
source= (Rank==0) ? NP-1:Rank-1;
\end{verbatim}
which means that the left neighbor of the process with the rank $0$ is the process $N_p-1$.

The received column of memory words is stored in the first column of the s-lattice which has been reserved for the border memory words of the neighbor process. When the border memory words are received, the left half of sub-lattice is updated (Fig.\ref{fig.updating}.c). Likewise, the destination and the source, in phase 2, are determined by the following code (Fig.\ref{fig.updating}.d):
\begin{verbatim}
destination = (Rank == 0) ? NP-1: Rank-1;
source= (Rank==NP-1) ? 0:Rank+1;
\end{verbatim}
The received column of the memory words is stored in the last column of the S-lattice which has been reserved for the border memory words of the neighbor process (Fig.\ref{fig.updating}.e).

\subsection{Calculating the Energy Of a Monte Carlo Iteration} \label{sub.sect.Energy}
In order to obtain the total energy of the lattice, the energy of all nearest neighbor pairs must be considered. However, if we consider the interaction energy of the right and down neighbors of each memory word, the total energy is calculated. Each process calculates the energy of each memory word in the S-lattice except for the first and last columns. Notice that the last column of the S-lattice contains the copy of the border memory words of the right process (Fig.~\ref{fig.updating}.e). Since these border memory words are not used until they are sent, the copy of them is still valid. This copied column is used as the right neighbor of the last column of the sub-lattice. The code in Listing.~\ref{listing.computeEnergy} shows how the interaction energy between a specific memory word and its right and down memory words is calculated:
\begin{lstlisting}[float,frame=trBL,breaklines=true,language=C++,caption=Computing the energy of a memory word,numbers=left,basicstyle=\footnotesize,label=listing.computeEnergy]
inline double computeEnergy(long int memoryWord,long int right,long int down)
{
    long int E = (memoryWord ^ right) + (memoryWord ^ down);
    double rv=0;
    for (int i = 1; i <= 21; i++)
    {
        switch (E & 7)
        {
        case 0:
            rv-=2*J;
            break;
        case 2:
            rv+=2*J;
            break;
        }
        E >>= 3;
    }
    return rv;
}
\end{lstlisting}
In the line 3, the outcome of the expression on the right side of the assignment operator, is a memory word which includes 21 3bit-groups. Every group contains a number between 0 and 2 i.e. 0 represents $-2J$, $1$ represents $0$ and $2$ represents $+2J$. Each 3bit-group retains the sum of the interaction energy between a specific spin with its right and down neighbors. The $\it for$ loop iterates on the 3bit-groups of $E$. In each iteration, the energy of one 3bit-group is extracted and is added to rv where rv retains the sum of energies of the 3bit-groups. Therefore, rv contains the total energy of the 21 3bit-groups.

\section{Results} \label{sect.result}
We have executed our program on a part of the computer cluster of Plasma Physics Research Center which includes 16 nodes networked by a switch. Each node is equipped with two Intel Xeon X5365 CPUs. We have used up to 9 nodes to test our program. Three different cases with different number of iterations have been considered in Tab.~\ref{table.cases}.
\begin{table}
\centering
  \begin{tabular}{|c|c|c|c|}
    \hline
    Test cases & Number of iterations & N & Average number of updates per spin in an iteration  \\
    \hline
    \hline
    1 & 5000 & 96 & 10  \\
    \hline
    2 & 4500 & 96 & 10  \\
    \hline
    3 & 4500 & 48 & 10  \\
    \hline
  \end{tabular}
  \caption{Three different cases which have been examined in this paper. The number of iterations, N and the average number of updates per spin in one iteration have been presented in the second, third and forth columns, respectively.}\label{table.cases}
\end{table}

\begin{figure}
  \centering
  \includegraphics[width=0.7\linewidth]{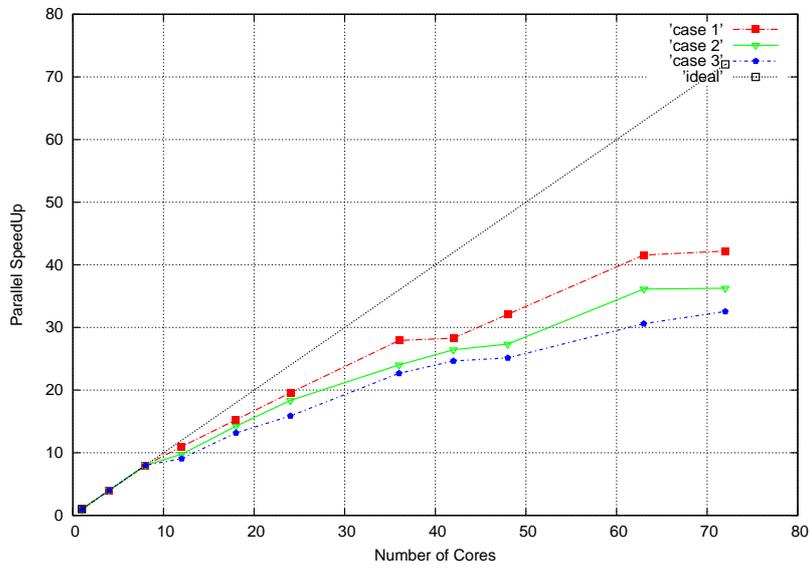}
  \caption{Parallel speed-up versus the number of cores presented in Tab.~\ref{table.cases}. }\label{fig.su}
\end{figure}

\begin{figure}
  \centering
  \includegraphics[width=0.7\linewidth]{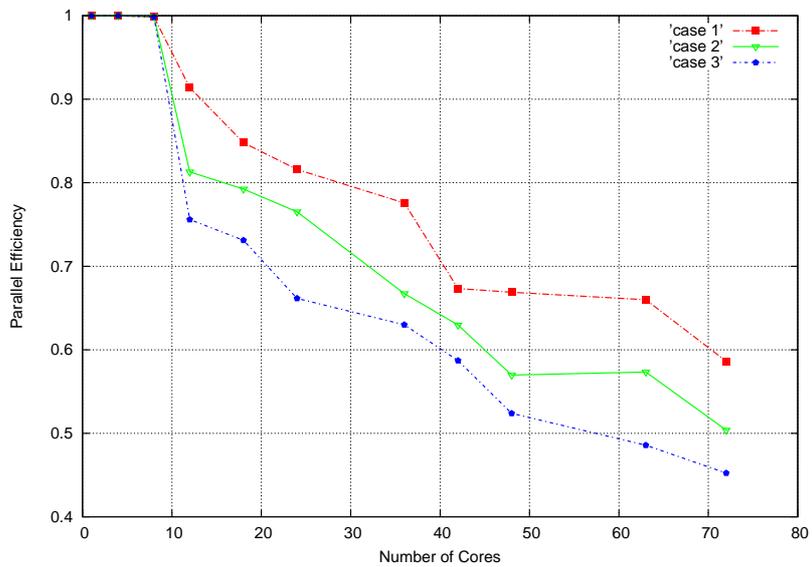}
  \caption{Parallel efficiency versus the number of cores presented in Tab.~\ref{table.cases}. }\label{fig.eff}
\end{figure}

The measured speed up and efficiency versus the number of cores are illustrated in Figs.~\ref{fig.su} and \ref{fig.eff}, respectively. As shown, for all three test cases, as number of cores and nodes is increased, efficiency goes down, especially when one more node is exploited, the efficiency drops considerably. This fall is due to the fact that overhead of the communication between processes on different nodes is higher than overhead of the communication between processes on the same node.

Now, we are able to inspect the impact of the lattice dimension and the number of Monte Carlo iterations on the performance of our algorithm. Comparing the test cases 2 and 3, it is inferred that bigger lattice sizes get better speedup and efficiency. In addition, the comparison between the test cases 1 and 2, we can claim when the number of Monte Carlo iterations increases, better speed up and efficiency is deduced. Therefore, our algorithm has better performance for bigger lattice sizes and higher number of Monte Carlo iterations.

\appendix
\section{Source Code} \label{Appendix.Code}
\begin{cframed}[white]{
\begin{lstlisting}[frame=lines,breaklines=true,language=C++,caption=Source Code,numbers=left,basicstyle=\ttfamily \footnotesize,label=listing.source,commentstyle=\color{blue}\scriptsize\textit, keywordstyle=\color{magenta}\textbf,stringstyle=\color{orange}]
#include <iostream>
#include <math.h>
#include <random>
#include <bitset>
#include <fstream>
#include <mpi.h>
using namespace std;
const double J=1; // coupling coefficient
const double K=1; // Boltzmann Coefficient
ranlux48 &engine(bool changeSeed)
{
    static random_device seed;
    static ranlux48 e(seed());
    if(changeSeed)
        e.seed(seed());
    return e;
}
inline double realRand()
{
    static uniform_real_distribution<double> d {0,1};
    return d(engine(false));
}
inline int intRand()
{
    static uniform_int_distribution<int> d {0,1};
    return d(engine(false));
}
void update(long int &memoryWord,long int up,long int right,long int down,long int left,double Temp)
{
    double coeff=-4*J/(K*Temp);
    long int E = (memoryWord ^ right) + (memoryWord ^ left) + (memoryWord ^ up) + (memoryWord ^ down);
    long int mustFlipSpins=0; // the memory word containing Spins to be flipped
    for(int i=0; i<21; i++)
    {
    	//interpretaion of 3bit-groups and decide what to do
        switch(E & 7)
        {
        case  0:    //if deltaE=+8J
            if (realRand()< exp(2*coeff))
                mustFlipSpins |=(1L << (3*i));
            break;
        case 1:     //if deltaE=+4J
            if (realRand()< exp(coeff))
                mustFlipSpins |=(1L << (3*i));
            break;
        default:   //if deltaE= 0 or -4J or -8J
            mustFlipSpins |=(1L << (3*i));
            break;
        }
        E >>= 3;
    }
    memoryWord ^= mustFlipSpins; // flipping spins
}
inline void downBoundaryCondition(long int &down,long int upperMemoryWord)
{
    down = upperMemoryWord;
    down  <<= 4;
    down  >>= 1;
    down |= (upperMemoryWord >> 60);
}
inline void upBoundaryCondition(long int &up,long int downMemoryWord)
{
    up = downMemoryWord;
    up  >>= 3;
    up |= ((downMemoryWord & 7) << 60);
}
inline double computeEnergy(long int memoryWord,long int right,long int down)
{
    long int E = (memoryWord ^ right) + (memoryWord ^ down);
    double rv=0;
    for (int i = 1; i <= 21; i++)
    {
        switch (E & 7)
        {
        case 0:
            rv-=2*J;
            break;
        case 2:
            rv+=2*J;
            break;
        }
        E >>= 3;
    }
    return rv;
}
int main()
{
    //initialize the MPI Library
    MPI_Init(NULL, NULL);
    //Rank of Process
    int Rank;
    MPI_Comm_rank(MPI_COMM_WORLD, & Rank);
    // Number Of Processes
    int NP;
    MPI_Comm_size(MPI_COMM_WORLD, & NP);
    // size of lattice is N x 21N
    const int N = 4;
    //Temperature range and rise-unit
    double Temp = 0.001,TempMax=10.001,dT=0.1;
    // No of Columns dedicated to each individual process
    const int NC = 21 * N / NP;
    // No of Monte Carlo Iterations
    int noOfIterations = 400;
    //No of Memory Words in phase 1
    int noOfMemoryWordsP1 = N*floor(NC/2);
    //No of Memory Words in phase 2
    int noOfMemoryWordsP2 = (N*NC) - N*floor(NC/2);
    //No of Memory Words to be updated in phase 1
    int noOfUpdatesP1 = noOfMemoryWordsP1 *10;
    //No of Memory Words to be updated in phase 2
    int noOfUpdatesP2 = noOfMemoryWordsP2 *10;
    // 2D array to retain s-Lattice
    long int s_lattice[N][NC+2];
    //a random number generator to select a row from s-lattice
    uniform_int_distribution<int> rowGenerator(0,N-1);
    //a random number generator to select a column from s-lattice
    uniform_int_distribution<int> colGenerator(1,NC);
    /*---------s-lattice Initialization----------*/
    for(int i=0; i<N; i++)
    {
        for(int j=1; j<NC+1; j++)
        {
            //initialize memory word
            s_lattice[i][j]=0;

            for(int k=1 ; k<=21; k++)
            {
                s_lattice[i][j]<<=3;
                if(intRand()== 1)
                    s_lattice[i][j]++;
            }
        }
    }
    /*------------------------------*/
    //Loop over different levels of temperature
    while (Temp <= TempMax)
    {
        //Average Energy of SubLattice: sum of all energies in all iterations for a specific temperature
        double AES = 0;
        //Monte-Carlo Iterations
        for (int iter = 1; iter <= noOfIterations; iter++)
        {
            //variables(memory-words) to keep a memory-Word neighbors
            long int  right, left, up, down;
            //variables to keep destination and source of send and receive
            int destination=0,source=0;
            //businessBasket is used as container for interprocess communication. it contains a column of memory-words.
            long int businessBasket[N] {};
            /*------------Phase 1-------------------*/
            //reset random generators
            colGenerator.param(uniform_int_distribution<int>::param_type(1,floor(NC/2)));
            colGenerator.reset();
            rowGenerator.reset();
            engine(true);
            /*-----InterProcess Communication-------*/
            //destination to which the rightmost column will be send
            destination = (Rank == NP-1) ? 0: Rank+1;
            //source from which the column will be received
            source= (Rank==0) ? NP-1:Rank-1;
            //fill the basket with the rightmost column of sublattice
            for (int i = 0; i < N; i++)
                businessBasket[i]=s_lattice[i][NC];
            //send the basket to destination asynchronously
            MPI_Send(&businessBasket,N,MPI_LONG,destination,1,MPI_COMM_WORLD);
            //receive the column from source synchronously
            MPI_Recv(&businessBasket,N,MPI_LONG,source,1,MPI_COMM_WORLD,MPI_STATUS_IGNORE);
            //update s-lattice with respect to receiving data
            for (int i = 0; i < N; i++)
                s_lattice[i][0]=businessBasket[i];
            /*---------------------------------------*/
            //randomly select and update memory words of s-lattice
            for (int i = 0; i < noOfUpdatesP1; i++)
            {
                //select a memory word randomly
                int r = rowGenerator(engine(false));
                int c = colGenerator(engine(false));
                /*assigning neighbors of the memory-word*/
                right = s_lattice[r][c + 1];
                left = s_lattice[r][c - 1];
                //to assign down-neighbor,a memory-word in the last row of sub_lattice is treated differently
                if (r == N - 1)
                    downBoundaryCondition(down,s_lattice[0][c]);
                else
                    down = s_lattice[r + 1][c];
                //to assign up-neighbor,a memory-word in the first row of sub_lattice is treated differently
                if (r == 0)
                    upBoundaryCondition(up,s_lattice[N - 1][c]);
                else
                    up = s_lattice[r - 1][c];
                /*---------------------------------------*/
                //update selected memory-word
                update(s_lattice[r][c] ,up,right,down,left,Temp);
            }
            /*-----------End OF Phase 1----------*/
            /*------------Phase 2-------------*/
            //reset random generators
            colGenerator.reset();
            rowGenerator.reset();
            colGenerator.param(uniform_int_distribution<int>::param_type(1 + floor(NC/2),NC));
            engine(true);
            /*----InterProcess Communication-------*/
            //destination to which the column will be send
            destination = (Rank == 0) ? NP-1: Rank-1;
            //source from which the column will be received
            source= (Rank==NP-1) ? 0:Rank+1;
            //fill the basket with leftmost column of sub-lattice
            for (int i = 0; i < N; i++)
                businessBasket[i]=s_lattice[i][1];
            //send the basket to destination asynchronously
            MPI_Send(&businessBasket,N,MPI_LONG,destination,1,MPI_COMM_WORLD);
            //receive data from source synchornously.
            MPI_Recv(&businessBasket,N,MPI_LONG,source,1,MPI_COMM_WORLD,MPI_STATUS_IGNORE);
            //update s-lattice with respect to receiving data
            for (int i = 0; i < N; i++)
                s_lattice[i][NC+1]=businessBasket[i];
            /*-----------------------------------------------*/
            //randomly select and update memory words of sub-lattice
            for (int i = 0; i < noOfUpdatesP2; i++)
            {
                //select a memory word randomly
                int r = rowGenerator(engine(false));
                int c = colGenerator(engine(false));
                /*assigning neighbors of the memory-word*/
                right = s_lattice[r][c + 1];
                left = s_lattice[r][c - 1];
                //to assign down-neighbor,a memory-word in the last row of sub_lattice is treated differently
                if (r == N - 1)
                    downBoundaryCondition(down,s_lattice[0][c]);
                else
                    down = s_lattice[r + 1][c];
                //to assign up-neighbor,a memory-word in the first row of sub_lattice is treated differently
                if (r == 0)
                    upBoundaryCondition(up,s_lattice[N - 1][c]);
                else
                    up = s_lattice[r - 1][c];

                /*---------------------------------------*/
                //update selected memory-word
                update(s_lattice[r][c] ,up,right,down,left,Temp);
            }
            /*---------End OF Phase 2------------*/
            /*------Calculate Energy of Iteration------*/
            double E_Iteration = 0;
            //computing potential energy between each individual spin with its right and down-neighbor
            for (int i = 0; i < N; i++)
            {
                for (int j = 1; j < NC+1; j++)
                {
                    /*assigning neighbors of the memory-word*/
                    right = s_lattice[i][j + 1];
                    //to assign down-neighbor,a memory-word in the last row of sub_lattice is treated differently
                    if (i == N - 1)
                        downBoundaryCondition(down,s_lattice[0][j]);
                    else
                        down = s_lattice[i + 1][j];
                    /*--------------------------------------*/
                    //compute and add energy to E_Iteration
                    E_Iteration+=computeEnergy(s_lattice[i][j],right,down);
                }
            }
            /*----------------------------------------*/
            AES += E_Iteration;
        }
        //average on total iterations
        AES /= noOfIterations;
        //Map Reduce
        //the result will be sent to process 0
        double E_Global_lattice=0;
        MPI_Reduce( & AES, & E_Global_lattice, 1, MPI_DOUBLE, MPI_SUM, 0, MPI_COMM_WORLD);
        if (Rank == 0)
        {
            //sending the result to output
            cout << Temp << "\t" << E_Global_lattice/pow(21*N,2) << "\n";
        }
        //temperature raised to next level
        Temp += dT;
    }
    //Free resources of MPI
    MPI_Finalize();
    return 0;
}

\end{lstlisting}}
\end{cframed}

\bibliographystyle{unsrt}
\bibliography{article}

\end{document}